\documentclass[aps,pre,twocolumn,superscriptaddress,showpacs]{revtex4}
\usepackage{graphics,amsmath,amssymb}
\newcommand{\csse}{\affiliation{Department of Mathematics, Center
for Systems Science and Engineering Research, Arizona State
University, Tempe, Arizona 85287}}
\newcommand{\ee}{\affiliation{Department of Electrical Engineering,
Arizona State University, Tempe, Arizona 85287}}

\begin{document}

Phys. Rev. E {\bf 66}, 046139 (2002)
\title{Smallest small-world network}

\author{Takashi Nishikawa}
\email{tnishi@chaos6.la.asu.edu}
\csse
\author{Adilson E. Motter}
\email{motter@chaos3.la.asu.edu}
\csse
\author{Ying-Cheng Lai}
\csse
\ee
\author{Frank C. Hoppensteadt}
\csse
\ee
\begin{abstract}
  \vspace{5mm}Efficiency in passage times is an important issue in designing
  networks, such as transportation or computer networks.  The small-world
  networks have structures that yield high efficiency, while keeping
  the network highly clustered.  We show that among all networks with
  the small-world structure, the most efficient ones have a single
  ``center'', from which all shortcuts are connected to uniformly
  distributed nodes over the network. The networks with several centers
  and a connected subnetwork of shortcuts are shown to be ``almost'' as
  efficient.  Genetic-algorithm simulations further support our
  results.
\end{abstract}

\pacs{89.75.Hc, 45.10.Db, 89.20.Hh}

\maketitle


The small-world network models have received much attention from
researchers in various disciplines, since they were introduced by
Watts and Strogatz~\cite{watts1998} as models of real networks that
lie somewhere between being random and being regular.  Small-world
networks are characterized by two numbers: the average path length $L$
and the clustering coefficient $C$.  $L$, which measures
efficiency~\cite{footnote1} of
communication or passage time between nodes, is defined as being the
average number of links in the shortest path between a pair of nodes
in the network.  $C$ represents the degree of local order, and is
defined as being the probability that two nodes connected to a common
node are also connected to each other.

Many real networks are sparse in the sense that the number of links in
the network is much less than $N(N-1)/2$, the number of all possible
(bidirectional) links.  On one hand, random sparse networks have short
average path length (i.e., $L \sim \log N$), but they are poorly
clustered (i.e., $C \ll 1$).  On the other hand, regular sparse
networks are typically highly clustered, but $L$ is comparable to $N$.
(All-to-all networks have $C=1$ and $L=1$, so they are most efficient,
but most expensive in the sense that they have all $N(N-1)/2$ possible
connections and so they are dense rather than sparse.) The small-world
network models have advantages of both random and regular sparse
networks: they have small $L$ for fast communication between nodes,
and they have large $C$, ensuring sufficient redundancy for high fault
tolerance.  Many networks in the real world, such as the world-wide
web (WWW)~\cite{albert1999}, the neural network of
\emph{C.~elegans}~\cite{watts1998,watts1999}, collaboration networks of
actors~\cite{watts1998,watts1999}, networks of scientific
collaboration~\cite{newman2001}, and the metabolic network of
\emph{E.~coli}~\cite{wagner2001}, have been shown to have this property.
The models of small-world networks are constructed from a regular lattice
by adding a relatively small number of shortcuts at random, where a
link between two nodes $u$ and $v$ is called a \emph{shortcut} if the
shortest path length between $u$ and $v$ in the absence of the link is
more than two~\cite{watts1999}.  The regularity of the underlying
lattice ensures high clustering, while the shortcuts reduce the size
of $L$.

Most work has focused on average properties of such models over
different realizations of {\it random} shortcut configurations.
However, a different point of view is necessary when a network is to
be designed to optimize its performance with a restricted number of
long-range connections.  For example, a transportation network should
be designed to have the smallest $L$ possible, so as to maximize the
ability of the network to transport people efficiently, while keeping
a reasonable cost of building the network.  The same can be said about
communication networks for efficient exchange of information between
nodes.  We fix the number of shortcuts here and as a result the
clustering coefficient $C$ for any configuration of shortcuts is
approximately as high as that of the underlying lattice.  The problem
we address in this paper is: \emph{given a number of shortcuts in a
small-world network, which configuration of these shortcuts minimizes
$L$?}~\cite{footnote2}.

Most random choices of shortcuts result in a suboptimal configuration,
since they do not have any special structures or organizations.  On
the contrary, many real networks have highly structured configurations
of shortcuts.  For example, in long-range transportation networks, the
airline connections between major cities which can be regarded as
shortcuts, are far from being random, but they are organized around
hubs.  Efficient travel involves ground transportation to a nearest
airport, then flights through a hub to an airport closest to the
destination, and ground transportation again at the end.

\begin{figure}
\includegraphics{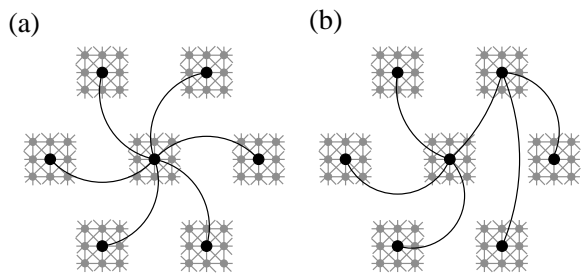}
\caption{Examples of shortcut configuration with (a) a single
center and (b) two centers.\label{fig1}}
\end{figure}

In the following, we show that the average path length $L$ of a
small-world network with a fixed number of shortcuts attains its
minimum value when there exists a ``center'' node, from which all
shortcuts are connected to uniformly distributed nodes in the
network~\cite{footnote3}.  An example of such a configuration is
illustrated in Fig.~\ref{fig1}(a).
We also show that if a small-world network has several ``centers'' and
its subnetwork of shortcuts is \emph{connected}, then $L$ is almost as
small as the minimum value.  An example of such configuration is shown
in Fig.~\ref{fig1}(b).  We then derive an explicit formula for the
minimum average path length in the case of the small-world network
models constructed from a one-dimensional lattice by adding a fixed
number of shortcuts.  Finally, we verify the results by performing
genetic-algorithm simulations for minimizing $L$.

\begin{figure}
\includegraphics{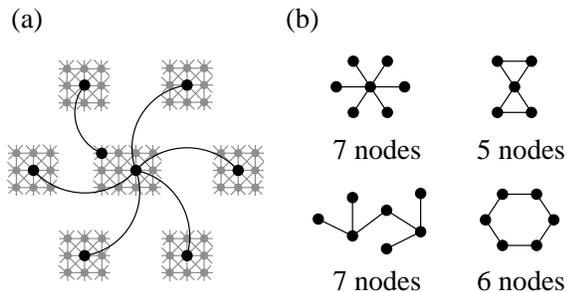}
\caption{(a) Configuration with one shortcut disconnected from
the rest of the subnetwork of shortcuts.  (b) Various
configuration of shortcuts with $m=6$ shortcuts.\label{fig2}}
\end{figure}

Our general argument proceeds as follows.  A small-world network is
composed of two parts: the underlying network (e.g., a regular
lattice) and the subnetwork of shortcuts containing only the shortcuts
and their nodes. Let $m$ denote the number of shortcuts.  First, for
$L$ to be as short as possible, the subnetwork of shortcuts must be
connected.  This connectivity is unlikely to happen if the shortcuts
are chosen at random, since the network is sparse.  Indeed, the
probability is less than $m!/N^{m-1}$, where $N$ is the number of
nodes in the network.  For example, for $N=1000$ and $m=10$, the
probability is smaller than $10^{-22}$.  Having a disconnected
component in the subnetwork of shortcuts increases the value of $L$.
In particular, consider the configuration of shortcuts as shown in
Fig.~\ref{fig2}(a), where one of the shortcuts in Fig.~\ref{fig1}(a)
is disconnected from the rest of the subnetwork of shortcuts.
If the shortest path between a pair of nodes involves going from the
disconnected shortcut to the rest of the subnetwork, then its length
is increased by 2 compared to the path length between the corresponding
pair in Fig.~\ref{fig1}(a).  This increases the average path length
$L$.

Next, observe that the nodes in the subnetwork of shortcuts must be
uniformly distributed over the network.  This can be seen by noting
that the average length of the shortest path from a node to its
nearest shortcut is smallest when these nodes are uniformly
distributed.

Finally, among all possible configurations of connected subnetworks of
shortcuts with uniformly distributed nodes, ones with a single center
involve the largest number of nodes (namely, $m+1$).
Figure~\ref{fig2}(b) shows some examples of connected subnetworks with
$m=6$.  Obviously, increasing the number of nodes involved in the
shortcut subnetwork reduces $L$, since it reduces the average path
length to the nearest shortcut node.  Among all connected
configurations of shortcuts having $m+1$ nodes, the ones having a
single center give the shortest value for $L$, since the average
path length of the shortcut subnetwork is the smallest in that case.

These arguments indicate that given a fixed number of shortcuts, the
networks with a connected subnetwork of shortcuts having nodes
uniformly distributed have smaller $L$ than a typical
random configuration, and among those the ones with a single center 
minimize $L$.  In other words, the ``smallest'' small-world
networks are characterized by these structures.

Now we will compute explicitly the average path length for a
configuration with a single center in the case of small-world networks
constructed from a one-dimensional lattice.  Consider $N$ nodes
arranged uniformly on a circle of unit circumference, where each node
is connected to its two nearest-neighbor nodes.  In addition, consider
shortcuts connecting $m$ arbitrary pairs of nodes.  To make the
calculation simple, we take the continuum limit $N \rightarrow \infty$
with $m$ fixed, in which the network becomes a continuous graph
composed of a circle corresponding to the lattice and chords
representing the shortcuts.  Let us define the distance $d(P,Q)$
between points $P$ and $Q$ on the continuous graph as the length of
the shortest continuous path along the graph, \emph{regarding the
length of a chord as zero}.  In other words, a shortcut is regarded as
identifying two points on the circle, rather than merely connecting
them.  Then, the number of links in the shortest path between nodes
$P$ and $Q$ in the original network, normalized by $N$, tends to
$d(P,Q)$ as $N \rightarrow \infty$.  This one-dimensional model,
despite being one of the simplest models of small-world networks,
captures basic features of many real networks.  In
Ref.~\cite{newman2000}, a mean-field-type argument was used to derive
an analytical expression for an average of $L$ over random
configurations of shortcuts, which was later improved in
Ref.~\cite{barbour}.  In the following, we derive an analytical
expression for the configuration with a single center.

\begin{figure}
\includegraphics{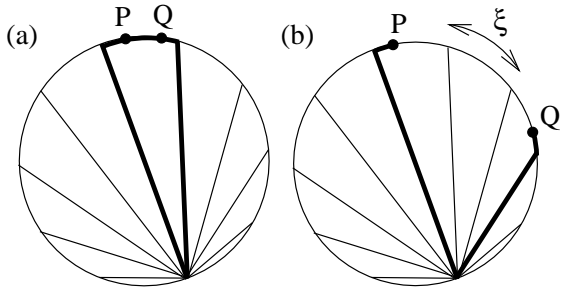}
\caption{The continuum limit model with configuration having a single center.
 (a) $Q$ is in $A_P$, the arc containing $P$, and (b) $Q$ is not in
 $A_P$. \label{fig3}}
\end{figure}

Consider the configuration of shortcuts with a center node connected
to $m$ other points on the circle, as shown in Fig.~\ref{fig3}.
The $m+1$ points including the center point are equally spaced with
$\xi \equiv 1/(m+1)$, and they divide the circle into $m+1$ arcs
of the same length.  We will compute the average $\overline{d(P,Q)}$
taken over all pairs $(P,Q)$.  Without loss of generality, we may
consider $P$ as fixed.  Let $A_P$ be the arc in which $P$ lies.
Suppose first that $Q \in A_P$ as in Fig.~\ref{fig3}(a).  Because the
end points of $A_P$ are connected to each other by two shortcuts via
the center, the distance in $A_P$ is equivalent to the distance on a
circle of circumference $\xi$.  Therefore, the average of $d(P,Q)$
over all pairs $(P,Q)$, such that $Q
\in A_P$, is equal to the average distance between two points on a
circle of circumference $\xi$, which is $\xi/4$.  Suppose now that $Q
\notin A_P$ as in Fig.~\ref{fig3}(b).  Let us denote the distance from
$P$ to its closest shortcut connection by $\alpha$, and the
distance from $Q$ to its closest shortcut by $\beta$.  Since the
shortest path between $P$ and $Q$ must pass through two shortcuts of
length zero, we have $d(P,Q) =
\alpha + \beta$.  Averaging this over all possible choices of $\alpha$
and $\beta$, which can take any value between 0 and $\xi/2$
independently, we obtain $\xi/2$.  Noting that the probabilities that $Q
\in A_P$ and that $Q \notin A_P$ are $1/(m+1)$ and
$m/(m+1)$, respectively, the normalized average path length $l$
can be calculated as
\begin{equation*}
l = \overline{d(P,Q)} = \frac{1}{m+1} \cdot \left( \frac{\xi}{4} \right) +
\frac{m}{m+1} \cdot \left( \frac{\xi}{2} \right) = \frac{2m+1}{4(m+1)^2}.
\end{equation*}

Let us now consider more general situation where each node in the
network has connections to its neighboring nodes, up to $k$th nearest
neighbors.  Because of the connections to $k$th nearest neighbors,
following the shortest path between nodes $P$ and $Q$ takes $1/k$
times less steps compared to the case discussed above.  Hence, we must
also scale $l$, the normalized average path length of the network, by
a factor $1/k$ yielding
\begin{equation}
l = \frac{1}{k}\overline{d(P,Q)} = \frac{2m+1}{4k(m+1)^2}.
\label{formula}
\end{equation}
An important observation about Eq.~(\ref{formula}) is that it can be
written as $l = f(m)/k$, where $f(m)$ is a function that depends only
on the number of shortcuts.  The formula derived in
Ref.~\cite{newman2000} for the average $l_r$ of normalized path
length over random configuration of shortcuts also has the same form
with different function for $f$, namely,
\begin{equation}
l_r =
\frac{1}{2k\sqrt{m^2+2m}}\tanh^{-1}\left(\frac{m}{\sqrt{m^2+2m}}\right).
\label{random}
\end{equation}
Note also that since the shortcuts are considered to have length zero,
the derivation above remains correct as long as the subnetwork of
shortcuts is connected and has uniformly distributed nodes, suggesting
that in the continuum limit these two conditions are sufficient to
achieve the minimum of $L$.

\begin{figure}
\includegraphics{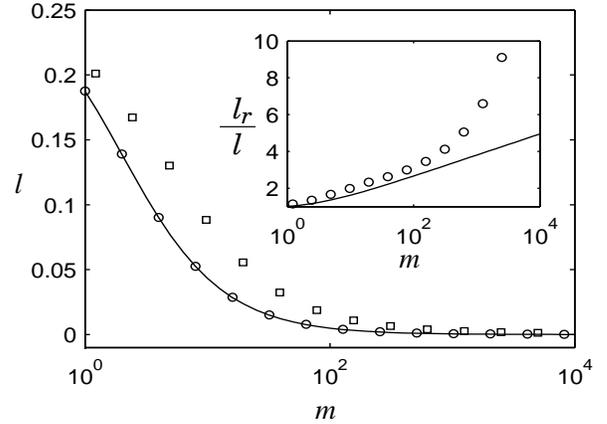}
\caption{Normalized path length of the network as a function 
of the number $m$ of shortcuts for $k=1$.  The continuous curve is
Eq.~(\ref{formula}). The circles and squares are the numerical
computation of $l$ for the configuration with a single center and of
$l_r$ over 10 random shortcut configurations, respectively.  The inset
shows the ratio $l_r/l$ computed from numerical simulations (circles)
and from theoretical results (\ref{formula}) and (\ref{random}) for
$N=\infty$ (continuous line).  $N=10^4$ was used for numerical
computations.
\label{fig4}}
\end{figure}

Figure~\ref{fig4} compares the calculation summarized in
Eq.~(\ref{formula}) (continuous curve) with numerical computation of
$l$ for a single center (circles) and of $l_r$ over 10 random
configurations of shortcuts (squares).
This shows an excellent agreement of Eq.~(\ref{formula}) with the
simulation.  In fact, the error in the Eq.~(\ref{formula}) due to the
approximation $N \rightarrow \infty$ is of order $1/N$, mainly because
the normalized length of a shortcut is considered to be zero rather
than $1/N$.  The inset in Fig.~\ref{fig4} shows the ratio $l_r/l$ as a
function of the number ($m$) of shortcuts.  Here the ratio is computed
from numerical simulations (circles) and from the theoretical results
~(\ref{formula}) and (\ref{random}) (continuous curve).  Since
Eq.~(\ref{formula}) is valid for $m \ll N$ and Eq.~(\ref{random}) is
valid for $1 \ll m \ll N$, the curve in the inset is exact in the
limit $N \rightarrow
\infty$ with $m\gg 1$ fixed.  Using the asymptotic
form $l_r \sim (\log 2m)/4m$ of Eq.~(\ref{random}) for $m \gg 1$, one
sees that $l_r/l \sim \log m$, explaining the fact that the curve in
the inset is almost a straight line for large $m$.  Numerical results
in the inset indicate that the effect of finite size and large
shortcut density actually increases the ratio, making the benefit of
optimizing the shortcut configuration to a single-center model even
larger than the theoretical prediction.

Finally, we simulate optimization of the shortcut configuration for a
one-dimensional array of nodes using the genetic-algorithm (GA)
methodology~\cite{mitchell}.  An initial population is described as
being a collection of various shortcut configurations specified by $m$
pairs of integers representing the locations of nodes connected by
shortcuts.  The fitness of each configuration is defined to be
$L^{-1}$, where $L$ is the average path length.  A new population of
shortcut configurations is created from the old one in analogy with
reproduction in population genetics: a configuration is viewed as
being the genome of an individual in the population, and in creating a
new population, we allow there to be one-point crossovers (i.e.,
interchanging subsets of shortcuts) and mutations
(i.e., changes in the location of end points by Gaussian random
numbers).  This creation process is continued until the fitness of the
best individual in the population is constant over 100 generations.
This gives a candidate for the optimal solution.  The program for the
simulation was developed using a C++ library called
GAlib~\cite{galib}.

\begin{figure}
\includegraphics{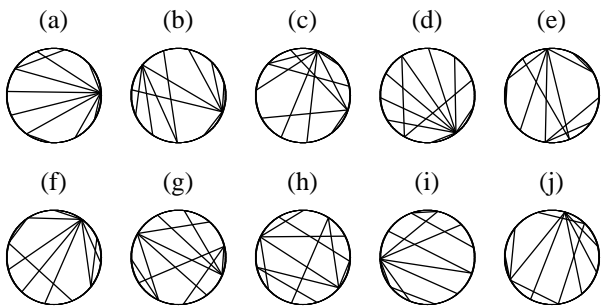}
\caption{Ten best solutions obtained by the genetic-algorithm
simulations. The corresponding average path lengths are (a) $L=44.962$
(b) $L=44.995$, (c) $L=45.043$, (d) $L=45.044$, (e) $L=45.163$, (f)
$L=45.221$, (g) $L=45.227$, (h) $L=45.275$, (i) $L=45.283$, (j)
$L=45.286$.  $N=1000$, $m=10$, and $k=1$ are used.\label{fig5}}
\end{figure}

Ten best solutions (here best means having shortest average path
length) resulting from 254 independent runs with $m=10$, $k=1$, and
$N=1000$, and the population size of 100 are shown in Fig.~\ref{fig5}.
First, observe that in each case the subnetwork of shortcuts is
connected.  This was the case in every solution found using the
genetic algorithm.  Second, in each case there are centers from which
many shortcuts emanate.  Moreover, the nodes in the subnetwork are
approximately equally spaced around the circle.  These observations
are consistent with the argument used above to establish our results.
All solutions in Fig.~\ref{fig5} have the average path length within
2\% of the average path length achieved by the single-center
configuration (which is 44.577).  In contrast, the corresponding value
for random shortcuts ($\approx 88$) is almost double the single-center
solution.  Although the single-center solution was not found by the
genetic algorithm due to the limited number (254) of simulation runs,
the results show that configurations with several centers are almost
as efficient as the single-center configuration, as long as the
subnetwork of shortcuts is connected and its nodes are uniformly
distributed.  The single-center solution was found for smaller
networks with $N=100$ and $m=5$, for which the computation is less
demanding.

\begin{figure}
\includegraphics{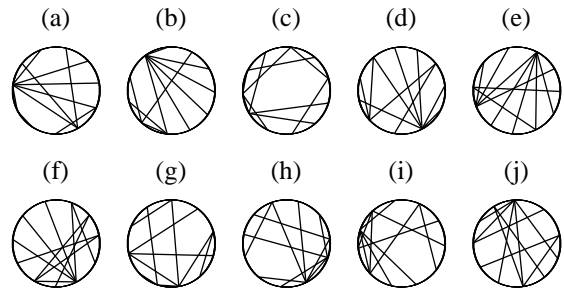}
\caption{Ten best solutions from 81 independent runs of GA simulation
with the population size of 30, $N=1000$, $m=10$, and $k=2$. The
corresponding average path lengths are (a) $L=24.309$, (b) $L=24.379$,
(c) $L=24.622$, (d) $L=24.627$, (e) $L=24.640$, (f) $L=24.650$, (g)
$L=24.653$, (h) $L=24.660$, (i) $L=24.779$, (j) $L=24.798$.  The
average path length is 23.795 for the single-center configuration,
while it is approximately 43 for random shortcuts. \label{fig6}}
\end{figure}

Any other values of $k$ should lead to similar results.  The case of
$k=2$ is shown in Fig.~\ref{fig6}.  In fact, due to the generality of
the argument given earlier, we expect that the results can be extended
to the case where the shortcuts are added to a lattice of higher
dimension, or to a regular network of another type.

The result of these simulations using the GA methodology shows that
design elements for efficient networks are (1) connectedness of the
shortcut subnetwork, (2) uniform distribution of nodes in the
subnetwork, and (3) existence of centers. 

We expect to see many examples of real networks with such structures.
Our computations on the neural network of \emph{C.~elegans} (which has
285 nodes, 2347 links, and 112 shortcuts) show that the structures are
indeed present: (i) the shortcut subnetwork has much fewer (= 15)
connected components than the average ($\approx$ 47) for randomly
chosen shortcuts, and the size of its giant component (= 75) is
significantly larger than the average ($\approx$ 12) over random
shortcuts; (ii) most ($\approx$ 88\%) of the nodes are within one step
of a shortcut; (iii) there are a few nodes having many shortcuts (11
shortcuts in the main center).  In general, a network with such
structures is robust against random failures, although it is sensitive
to deliberate attacks to the centers.  This property, which is shared
by scale-free networks~\cite{barabasi1999}, is shown to characterize
many real networks such as the Internet and the WWW~\cite{albert2000}.
However, some biological networks may be robust even against attacks
on the centers since loss of a center can result in shortcuts
reconnecting to nearby nodes followed by the optimization process that
quickly recovers the smallest configuration.

We have shown that among the small-world networks having a fixed
number of shortcuts, the average path length is smallest when there
exists a single center through which all of the shortcuts are
connected and shortcut nodes are uniformly distributed in the network.
We have also shown that the average path length is almost as small
when the shortcuts are connected and have a few centers, which was
supported by the result of the GA simulations.  Our results have
important consequences in situations where the efficiency of
information flow over a large network is required.  The fact that the
architecture of connected shortcuts with centers arises through
genetic algorithms suggests the possibility that such a structure could
emerge in networks in natural organisms (e.g., the neural network of
\emph{C.~elegans}), although the fitness used in GA here is not
necessarily related to that of natural selection in biology.  In
particular, it provides a potential mechanism for the appearance of
highly connected nodes while keeping high clustering in networks that
are evolving but not necessarily growing, such as neural and metabolic
networks.

T.N. was supported by DARPA/ONR through Grant N00014-01-1-0943.  A.E.M. thanks
Fapesp for financial support.  Y.C.L. was supported by AFOSR under Grant
No.~F49620-01-1-0317.  F.C.H. was supported in part by NSF through Grant
DMS-0109001.  We are grateful for helpful suggestions from
A.~P.~S.~de Moura.  


\end{document}